\documentclass[sn-mathphys, Numbered, iicol]{sn-jnl}
\usepackage{graphicx}%
\usepackage{multirow}%
\usepackage{amsmath,amssymb,amsfonts}%
\usepackage{amsthm}%
\usepackage{mathrsfs}%
\usepackage[title]{appendix}%
\usepackage{xcolor}%
\usepackage{textcomp}%
\usepackage{manyfoot}%
\usepackage{booktabs}%
\usepackage{algorithm}%
\usepackage{algorithmicx}%
\usepackage{algpseudocode}%
\usepackage{listings}%
\usepackage{cleveref}
\usepackage{ulem}
\usepackage{xspace}
\usepackage{mhchem}
\usepackage{subcaption}
\usepackage{bbm}
\usepackage{siunitx}
\makeatletter

\makeatletter
\newcommand*{\myfnsymbolsingle}[1]{%
  \ensuremath{%
    \ifcase#1% 0
    \or % 1
      \S%   
    \or % 2
      \dagger
    \or % 3  
      a
    \or % 4   
      b
    \or % 5
      c
    \or 
      d
    \else % >= 6
      \@ctrerr  
    \fi
  }%   
}   
\makeatother
\usepackage{alphalph}
\usepackage{fnpct} % just a recommendation, not needed.
\usepackage{xr}
\newcommand{\corres}[1]{}

\theoremstyle{thmstyleone}%
\theoremstyle{thmstyletwo}%

\theoremstyle{thmstylethree}%

\Crefname{figure}{\text{Fig.}}{\text{Figs.}}
\Crefname{section}{\text{Section}}{\text{Sections}}
\Crefname{equation}{\text{Eq.}}{\text{Eqs.}}

\usepackage{amsmath,amsfonts,bm}

\def\eqref#1{equation~\ref{#1}}
\def\1{\bm{1}}

\DeclareMathAlphabet{\mathsfit}{\encodingdefault}{\sfdefault}{m}{sl}
\SetMathAlphabet{\mathsfit}{bold}{\encodingdefault}{\sfdefault}{bx}{n}

\begin{document}
\setcounter{secnumdepth}{0}

\title[Article Title]{Mixture of Experts Architectures for Machine Learning Interatomic Potentials}

\author[1]{\fnm{Yuzhi} \sur{Liu}}

\author[1,2,3]{\fnm{Duo} \sur{Zhang}}

\author[1]{\fnm{Anyang} \sur{Peng}}

\author[1,3,4]{\fnm{Weinan} \sur{E}}

\author*[1,2]{\fnm{Linfeng} \sur{Zhang}}
\email{zhanglf@aisi.ac.cn}

\author*[5,6]{\fnm{Han} \sur{Wang}}
\email{wang\_han@iapcm.ac.cn}

\affil[1]{AI for Science Institute, Beijing 100080, P. R.~China}
\affil[2]{DP Technology, Beijing 100080, P. R.~China}
\affil[3]{Academy for Advanced Interdisciplinary Studies, Peking University, Beijing 100871, P. R.~China}
\affil[4]{
School of Mathematical Sciences, Peking University, Beijing 100871, P. R.~China
}
\affil[5]{National Key Laboratory of Computational Physics, Institute of Applied Physics and Computational Mathematics, Fenghao East Road 2, Beijing 100094, P.R.~China}
\affil[6]{HEDPS, CAPT, College of Engineering, Peking University, Beijing 100871, P.R.~China}

\abstract{
Machine Learning Interatomic Potentials (MLIPs) enable accurate large-scale 
atomistic simulations, yet improving their expressive capacity efficiently remains challenging. 
Here we systematically investigate Mixture-of-Experts (MoE) and Mixture-of-Linear-Experts (MoLE) 
architectures within the DPA3 framework for MLIPs and analyze the effects of routing strategies and expert designs. 
We show that sparse activation combined with shared experts yields substantial performance gains, 
and that nonlinear MoE formulations outperform MoLE when shared experts are present, 
underscoring the importance of nonlinear expert specialization. 
Furthermore, element-wise routing consistently surpasses configuration-level routing, 
while global MoE routing often leads to numerical instability. 
The resulting element-wise MoE model consistently outperforms all DPA3-based baselines 
across the OMol25, OMat24, and OC20M benchmarks.
Analysis of routing patterns reveals chemically interpretable expert specialization aligned with 
periodic-table trends, indicating that the model effectively captures element-specific 
chemical characteristics for precise interatomic modeling.
}

\maketitle

\small

\section{Introduction}\label{sec: Intro}

Over the past few years, MLIPs have rapidly advanced as a powerful strategy to bridge the 
divide between the high accuracy of Quantum Mechanical (QM) methods and the computational 
efficiency of classical force fields\cite{behler2007generalized, bartok2010gaussian, 
rupp2012fast, zhang2018deep, schutt2018schnet, gasteiger2020directional, batzner20223, 
deringer2019machine, unke2021machine, mishin2021machine,vcivzek1966correlation,white1992density,ceperley1980ground}. 
The fundamental principle of MLIPs lies in approximating the high-dimensional Potential Energy Surface(PES) 
by learning from extensive datasets generated via first-principles calculations, such as 
Density Functional Theory(DFT)~\cite{hohenberg1964inhomogeneous, kohn1965self, andolina2023highly,ceperley1980ground, perdew1996generalized,becke1993density}. 
Once trained, these potentials can be employed in large-scale molecular dynamics(MD) simulations, 
enabling the study of complex atomistic processes at a fraction of the computational cost of direct 
QM approaches~\cite{zhang2018deep, deringer2019machine, unke2021machine, mishin2021machine}. 
As a result, MLIPs have found successful applications across diverse fields, including drug 
discovery~\cite{badaoui2022combined, zeng2023qdpi,smith2017ani,rufa2020towards}, 
materials science~\cite{bartok2018machine, deringer2020general,wen2022deep,jia2020pushing,machaka2021machine}, 
and catalysis design~\cite{yang2022ammonia, car1985unified,chanussot2021open}, providing a scalable 
and accurate framework for exploring chemical space and predicting material properties.  

To more accurately model interatomic interactions, a variety of pioneering neural network architectures, 
such as  MACE~\cite{batatia2022mace}, SevenNet~\cite{park2024scalable}, DPA~\cite{zhang2024pretraining, zhang2024dpa}, 
eSEN~\cite{fu2025learning}, and AlphaNet~\cite{yin2025alphanet}, have been developed. 
In pursuit of broader applicability, researchers have progressively scaled up model parameters and trained 
these architectures on increasingly large datasets, often employing multi-task learning strategies~\cite{zhang2024dpa, wood2025family, shoghi2023molecules}.
A notable observation emerging from these efforts is the presence of scaling laws: similar to trends 
observed in Large Language Models (LLMs)~\cite{kaplan2020scaling, hoffmann2022training}, 
the performance of these interatomic potentials improves predictably with increases in 
both model size and training data volume~\cite{zhang2025graph, wood2025family}. 

This scaling behavior provides a clear and encouraging roadmap for advancing the accuracy and capability of future machine-learned potentials.

Despite the clear benefits of scaling, simply expanding model capacity through deeper or 
wider dense architectures remains computationally prohibitive in practice. 
Two primary bottlenecks hinder this approach: 
first, the all-to-all computational dependency in dense models limits parallel 
efficiency~\cite{narayanan2021efficient, shoeybi2019megatron}; 
second, as depth and width increase, the optimization landscape grows increasingly complex, 
resulting in diminishing returns and heightened training instability~\cite{kaplan2020scaling, hoffmann2022training}. 

MOE models have a long history in machine learning, originally 
introduced as ensemble methods in which multiple specialized models are 
combined under a gating network~\cite{jacobs1991adaptive, jordan1994hierarchical}. 
In modern deep learning, the sparsely-gated MoE layer, introduced by Shazeer 
et al.~\cite{shazeer2017outrageously}, revitalized this idea by replacing dense 
feed-forward networks with conditionally activated expert subnetworks. 
Subsequent work scaled MoE to massive language models through architectures 
such as GShard~\cite{lepikhin2020gshard}, Switch 
Transformer~\cite{fedus2022switch}, ST-MoE~\cite{zoph2022st}, 
and GLaM~\cite{du2022glam}, establishing MoE as a cornerstone 
of large-scale modeling~\cite{cai2025survey}.

An MoE layer replaces the standard dense feed-forward network 
with a collection of independent experts, governed by a 
learnable routing mechanism that activates only a sparse subset of parameters for any given input. 
This design effectively decouples the total model capacity 
from the computational cost per token~\cite{shazeer2017outrageously, lepikhin2020gshard}. 
Architecturally, MoE is inherently more conducive to parallelism; 
by distributing experts across different devices, it reduces 
the communication bottlenecks typical of standard model parallelism. 
Consequently, for a fixed computational budget, an MoE model can 
leverage a significantly larger total parameter count to achieve 
lower validation loss and superior representational power compared 
to its dense counterparts~\cite{fedus2022switch, zoph2022st, du2022glam, artetxe2021efficient}.

Although MoE offers a promising approach for improving the capacity 
of MLIPs, its direct adoption has not been fully realized in this domain.
Related ideas, such as divide-and-conquer, have been explored 
in interatomic potentials, where multiple specialized models are 
combined to improve accuracy and transferability~\cite{zeni2024divide}. 
While these approaches share the intuition of expert specialization, 
they rely on predefined partitions or hand-crafted weighting schemes, 
and therefore lack the flexibility of end-to-end learned routing 
and sparse activation mechanisms in modern MoE architectures.

More recently, the MoLE framework, exemplified by the UMA models, 
has been proposed to introduce expert-based modeling into MLIPs~\cite{wood2025family}. 
MoLE employs a linear combination of experts, ensuring compatibility with equivariant representations. 
It also constrains expert routing to depend only on configuration-level 
features, shared across all atoms, thereby maintaining smooth and differentiable potential energy surfaces.

However, these design choices fundamentally differ from standard 
MoE architectures with learned routing and sparse expert activation. 
In particular, the absence of nonlinear transformations on expert outputs may limit 
representational capacity, and the use of global, configuration-level 
gating precludes element-wise expert specialization within a single configuration.

As a result, the direct integration of such MoE architectures into MLIPs 
still faces several fundamental challenges. 
First, although MoE is not fundamentally incompatible with equivariant 
representations, applying conventional nonlinear dense MoE layers 
directly to equivariant tensor features is nontrivial, because both the expert transformations and the 
routing weights must preserve the required symmetry constraints. In particular, 
the experts should be equivariant maps and the routing weights should be invariant scalars.
Standard MoE mechanisms therefore cannot be directly integrated into MLIP architectures based 
on equivariant graph neural networks (GNNs), such as MACE~\cite{batatia2022mace}, SevenNet~\cite{park2024scalable}, 
eSEN~\cite{fu2025learning}, and AlphaNet~\cite{yin2025alphanet}, which rely on equivariant representations.
Second, unlike the discrete token representations in language models, 
where sparse expert activation is natural, MLIPs model continuous potential energy surfaces.
Abrupt expert switching in this setting can introduce numerical instabilities or non-physical discontinuities.

In this work, we propose an MoE-integrated DPA3 MLIP architecture.
DPA3 is a graph neural network built on a line-graph series and has 
demonstrated state-of-the-art accuracy across diverse benchmarks~\cite{zhang2025graph}, 
making it a robust baseline for evaluating MoE-based extensions.
Crucially, DPA3 employs exclusively invariant node and edge features, which permits 
valid nonlinear operations on expert outputs and thus enables the seamless 
incorporation of standard MoE mechanisms within its framework.
Since invariant GNNs can be viewed as a special case of equivariant GNNs, 
the MoLE architecture can also be implemented within DPA3, providing a 
unified and controlled setting for direct comparisons between MoE and 
MoLE designs in terms of their effectiveness for improving MLIP performance.
Building on this foundation, we introduce element-wise expert gating, yielding 
an element-dependent MoE that models potential-energy contributions through species-specific expert routing.
In addition, we incorporate a shared-expert mechanism~\cite{dai2024deepseekmoe}, 
ensuring that a subset of experts is always activated to capture knowledge common across all element types.

The effectiveness and robustness of the proposed MoE design are rigorously validated 
on the OMol25, OMat24, and OC20M benchmarks, where it achieves substantial improvements over the baseline.
Notably, Principal Component Analysis (PCA) of the expert latent representations 
reveals that the learned expert distributions spontaneously recover trends consistent 
with the periodic table, suggesting that the model implicitly encodes 
fundamental chemical identities to achieve enhanced predictive accuracy.

\section{Results}\label{sec: res}

\subsection{Architectural overview and benchmark setup}

\begin{figure*}
    \centering
    \includegraphics[width=1.0\linewidth]{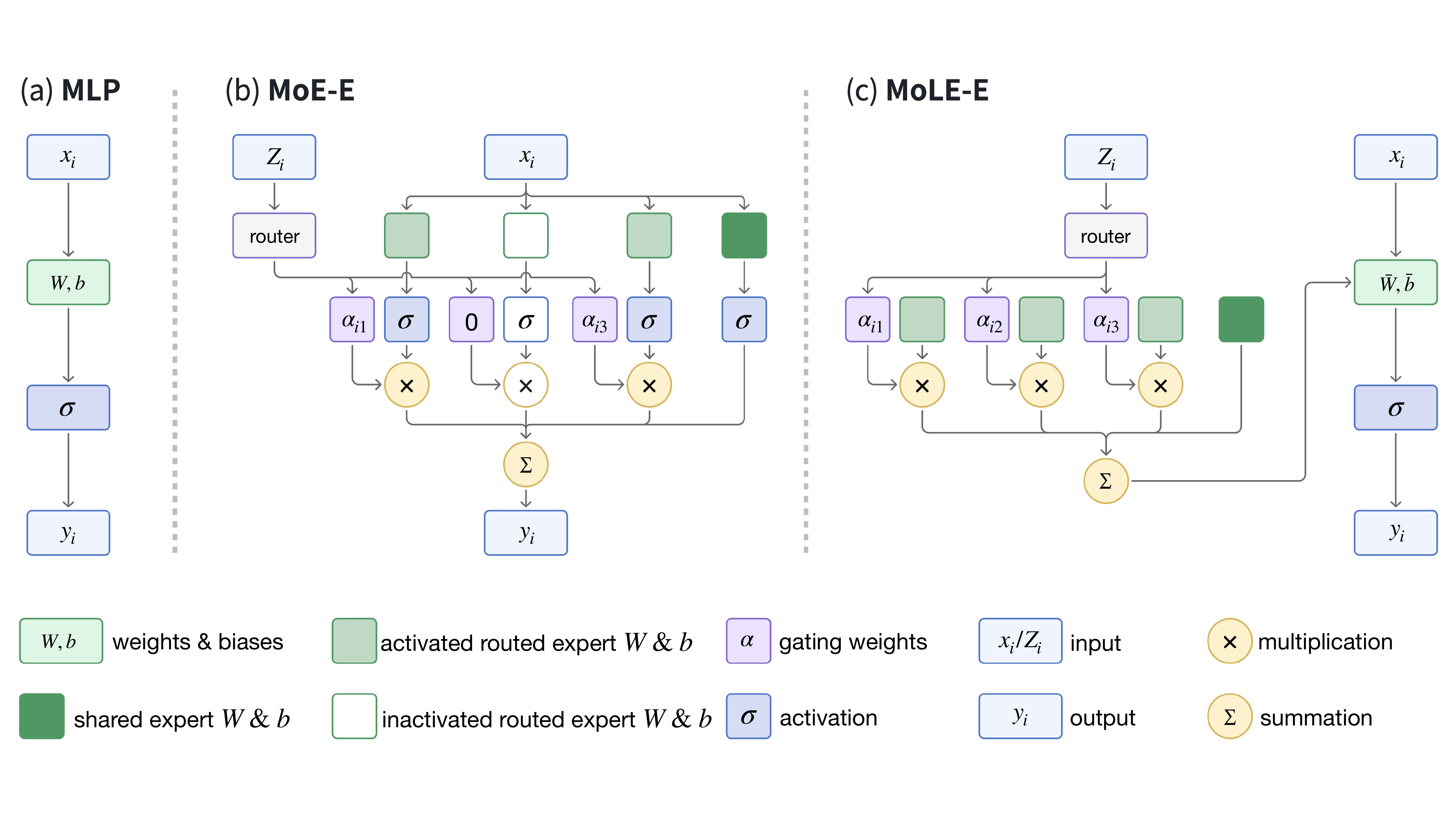}
    \caption{Schematic illustration of the model architectures. (a) The standard MLP framework. 
    (b) The MoE-E framework, featuring a router that dynamically selects 3 active experts from the pool, 
    complemented by 1 fixed shared expert to capture universal features. 
    (c) The MoLE-E framework, where 4 expert weights are linearly combined via the router to modulate the main network weights, 
    including 1 shared expert component.
    }
    \label{fig:MoE}
\end{figure*}

  We build on the DPA3 graph neural network architecture~\cite{zhang2025graph},
  in which most trainable parameters reside in dense feed-forward blocks used
  during message passing. In the baseline dense model, each such block performs
  a single nonlinear transformation of an atom-wise feature, as illustrated in
  Fig.~\ref{fig:MoE}(a). To increase capacity while controlling activation cost,
  we replace these dense blocks with MoE modules.

  As shown in Fig.~\ref{fig:MoE}(b), an MoE module contains $N$ parallel expert
  subnetworks, each of which is itself a relatively small dense layer. A router
  assigns gating weights to the experts based on the chemical identity of atom
  $i$, encoded by its atomic number $Z_i$. Only a subset of experts is activated
  for a given atom (sparse activation), while a fixed subset, called
  shared experts, is always active and captures features common to all
  chemical species. We consider two routing strategies. In element-wise
  routing, the router acts on a per-atom embedding $u_i$ derived from $Z_i$; we
  refer to this variant as \emph{MoE-E}. In global routing, following the
  atom-averaged scheme used in UMA~\cite{wood2025family}, the same gating weights
  are shared across all atoms in a configuration through mean pooling,
  \begin{equation}
  \bar{u} = \frac{1}{N_{\mathrm{atom}}} \sum_i u_i,
  \end{equation}
  and we denote this variant as \emph{MoE-G}.

  In addition to the standard MoE, we consider a linear-expert variant, MoLE,
  illustrated in Fig.~\ref{fig:MoE}(c). Whereas MoE first applies a nonlinear
  expert transformation and then combines the outputs, MoLE first linearly
  combines the expert weights and biases via the gating mechanism and then
  applies a single nonlinearity. The corresponding element-wise and global
  routing variants are denoted \emph{MoLE-E} and \emph{MoLE-G}, respectively. 
  Detailed mathematical formulations of MoE-E and MoLE-E are provided in Methods; 
  their global-routing counterparts are obtained by replacing $u_i$ 
  with the configuration-averaged vector $\bar{u}$.

  To evaluate these architectures, we adopt a standard 6-layer DPA3 model as
  the baseline, with hyperparameters given in Supplementary Table~1. Most
  experiments are conducted with 6-layer or 12-layer models on a single GPU,
  rather than the 24-layer multi-GPU setting of the original DPA3 work, because
  our focus is on systematically comparing architectural design choices under a
  computationally accessible setting.

  For the ablation studies, we report normalized Mean Absolute Error (MAE),
  defined as the ratio between the validation MAE of a given model and that of
  the 6-layer dense DPA3 baseline trained under the same conditions. Since all
  MoE and MoLE variants are constructed by replacing the corresponding
  feed-forward blocks with expert modules, this baseline serves as a natural
  reference; values below 1.0 indicate improved accuracy. The normalized metric
  is used only for controlled comparisons within the ablations, and we interpret
  the results together across energy and force metrics rather than relying on a
  single value. Absolute MAEs are provided in Supplementary Note~2. For the
  multi-dataset benchmarks, we report absolute prediction errors directly: MAE
  for energy and forces across all datasets, and additionally RMSE for OC20M to
  characterize the contribution of larger-error cases.

  We evaluate 6-layer variants of MoE and MoLE with both element-wise and global
  routing, and systematically compare their performance against the baseline to
  quantify the gains enabled by different expert designs.

\subsection{Sparse activation and shared experts}

\begin{figure*}
    \centering
    \includegraphics[width=0.9\linewidth]{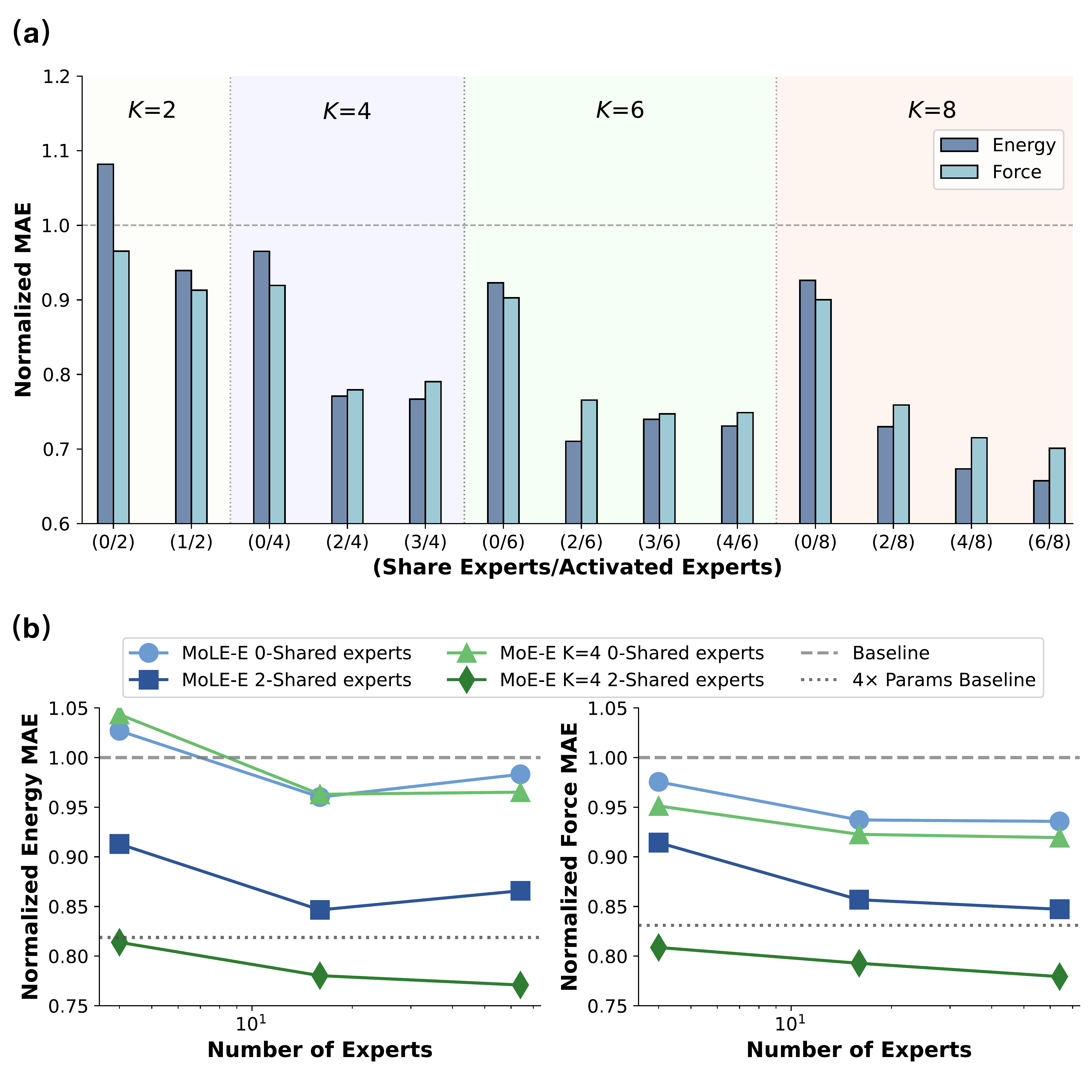}
    \caption{Performance benchmarks on the OMol25 dataset for the MoE-E 
    model with a total pool of $N=64$ experts. (a) Impact of 
    shared expert allocation. Normalized energy and force MAE are 
    shown as a function of the shared expert ratio. The x-axis denotes 
    the proportion of shared experts (e.g., $(2/6)$ indicates 2 shared 
    experts among 6 total activated experts). Background shading demarcates 
    different $K$ activation settings for MoE-E. The horizontal dashed line 
    provides a performance baseline corresponding to a standard 6-layer DPA3 
    model (approximately 1.3~M parameters). See Supplementary Table 2 for detailed values. 
    (b) Effect of total expert count: Comparative analysis of normalized energy (left) 
    and force (right) MAE as a function of the total number of experts. Two baselines are provided: 
    the standard 6-layer DPA3 model (approximately 1.3~M parameters) and a ``\(4\times\) Params'' 
    variant (approximately 5.2~M parameters) achieved by doubling the hidden 
    dimension of the 6-layer DPA3 model. See Supplementary Table 3 for detailed values.}
    
    \label{fig:share_experts}
\end{figure*}

Two important design choices in MoE are \textbf{sparse expert activation} and the use of \textbf{shared experts}.
Fig.~\ref{fig:share_experts}(a) systematically investigates the impact of these design choices using the OMol25 dataset as a benchmark~\cite{levine2025open}.
Covering 83 chemical elements, OMol25 displays remarkable chemical diversity, including complex intermolecular interactions, explicit solvation, variable charge and spin states, as well as reactive configurations.
In this experiment, the total number of experts is fixed at \(N = 64\), while the number of activated experts \(K\) varies from 2 to 8.
For each value of \(K\), multiple configurations with different numbers of shared experts are evaluated.
The specific values of \(N\) and \(K\) are chosen with future distributed expert parallelism in mind. 
A typical GPU compute node is equipped with 8 cards, and selecting \(N=64\) (a multiple of 8) ensures that experts can be evenly distributed across devices in an expert-parallel implementation. 
Similarly, the activated expert counts \(K=2, 4, 6, 8\) are even values that facilitate balanced load assignment across the activated experts. 
While the current ablation experiments are conducted on a single GPU, these choices are intended to be compatible with large-scale multi-GPU training and inference.

Overall, model performance improves with increasing \(K\).
In the absence of shared experts, the normalized energy (force) MAEs for \(K=2, 4, 6,\) and \(8\) are
1.082 (0.965), 0.965 (0.919), 0.923 (0.902), and 0.926 (0.900), respectively.
This trend is primarily driven by the effective linear expansion of model capacity, since a 
configuration with \(K\) activated experts corresponds to an approximately \(K\)-fold increase in the number of active parameters.
However, several nontrivial behaviors are observed.
Notably, the \(K=2\) model without shared experts, despite having roughly twice the parameter count of the baseline, exhibits a degraded energy MAE.
In addition, performance saturates between \(K=6\) and \(K=8\) under the tested conditions, 
indicating diminishing returns where the additional computational budget no longer translates into measurable accuracy gains.
These results suggest that a naive increase in the number of activated experts alone is insufficient.

Across nearly all choices of activated experts \(K\), introducing shared experts consistently improves model performance.
This effect is particularly pronounced for \(K=8\), where increasing the number of shared 
experts from zero to six reduces the normalized MAEs for energy and force by 0.269 and 0.200, respectively.
Notably, within the tested 6-layer DPA3 configurations on OMol25, performance is optimized 
when the number of shared experts approaches approximately half of the activated experts, as observed for the \(K=4, 6,\) and \(8\) configurations.
Beyond this regime, further increasing the fraction of shared experts yields diminishing 
returns or even slight performance degradation, for example, the force MAE increases when the number of shared experts rises from 2 to 3 in the \(K=4\) configuration.
Overall, these results indicate that, under the tested settings, allocating roughly half 
of the activated experts as shared experts provides an effective trade-off between capturing 
global chemical knowledge common across elements and preserving sufficient capacity for element-specific specialization.

When adopting a configuration in which half of the activated experts are shared, the normalized energy (force) MAEs for
\(K = 2, 4, 6,\) and \(8\) are reduced to 0.939 (0.913), 0.771 (0.779), 0.740 (0.747), and 0.673 (0.715), respectively.
In stark contrast to the no–shared-expert setting, the accuracy improvement in this regime 
does not plateau at \(K=8\), but instead continues to scale favorably with increasing model capacity.
These results highlight that shared experts are indispensable for effectively exploiting 
larger expert pools and for achieving consistent, monotonic accuracy improvements with increasing activated capacity.
Importantly, for a fixed number of activated experts \(K\), increasing the number 
of shared experts does not alter the total parameter count or computational cost.
Consequently, the observed accuracy gains from shared experts are achieved at essentially no additional computational overhead.

\subsection{MoE-E vs MoLE-E}
Fig.~\ref{fig:share_experts}(b) presents a comparison 
between the MoE-E and MoLE-E architectures, both of which maintain an identical parameter count. 
The fundamental distinction between these two 
designs lies in the sequencing of the nonlinear transformation relative to expert aggregation: 
MoE-E applies the activation function locally within each expert before mixing their outputs, 
whereas MoLE-E performs a linear aggregation of expert 
contributions followed by a global activation function.
Furthermore, while MoE-E utilizes a sparse activation 
of $K$ experts, MoLE-E employs a dense approach where all experts are activated.
For a controlled comparison, all MoE-E configurations 
are evaluated with the same number of activated experts (\(K=4\)).
The normalized energy and force MAEs of the MoE-E 
and MoLE-E architectures are reported as functions of the total number of experts \(N\). 

This setup also provides an ablation for separating nonlinear expert decomposition from sparse expert selection.
When \(N=4\) and \(K=4\), all experts in MoE-E are activated in every forward pass.
This configuration therefore behaves as a dense nonlinear MoE without sparse selection from a larger expert pool.
Increasing \(N\) to 16 or 64 while keeping \(K=4\) fixed introduces sparse expert selection under the same activated-expert budget, allowing us to assess whether selecting from a larger pool provides additional benefits beyond dense nonlinear expert decomposition.

Without shared experts, increasing \(N\) from 4 to 16 improves both MoE-E and MoLE-E, while the gains saturate when \(N\) is further increased to 64.
For MoE-E, this indicates that sparse selection from a larger expert pool can improve over the dense \(N=K=4\) setting, but that the benefit is limited when shared experts are absent.
Across the tested expert scales, MoE-E and MoLE-E show similar normalized MAEs without shared experts, suggesting that nonlinear expert decomposition alone is not sufficient to produce a large advantage over the linear MoLE formulation.

The integration of shared experts yields markedly different gains for the two architectures.
When two shared experts are introduced, the MoLE-E variant reduces the energy MAE by \(11.1\%\text{--}11.9\%\) 
and the force MAE by \(6.3\%\text{--}9.5\%\) relative to its no-shared-expert counterpart across
\(N=4,16,\) and \(64\).
Although the force MAE of MoLE-E continues to decrease slightly when the total number of experts 
increases from \(N=16\) to \(N=64\), the corresponding energy MAE increases by approximately \(2.3\%\), partially
offsetting the overall benefit of using a larger expert pool.

In contrast, MoE-E shows a stronger response to the introduction of shared experts.
For the same transition from zero to two shared experts, MoE-E reduces the energy MAE by \(19.0\%\text{--}22.0\%\) 
and the force MAE by \(14.1\%\text{--}15.2\%\) across the same range of expert counts.
Moreover, with two shared experts, increasing the total number of experts from \(N=16\) to \(N=64\) 
further reduces the energy and force MAEs of MoE-E by approximately \(1.2\%\) and \(1.7\%\), respectively.
These results indicate that shared experts are more effective in the nonlinear MoE-E architecture and 
that MoE-E can better leverage a larger expert pool under the tested settings.

\subsection{Parameter Efficiency of MoE-E}

Since the MoE-E model with four activated experts (\(K=4\)) activates approximately four times as 
many parameters as the standard 6-layer DPA3 baseline, we construct a widened dense baseline for comparison
under a similar activated-parameter budget.
This ``\(4\times\) Params Baseline'' is obtained by doubling the hidden feature dimension, which 
increases the dominant dense-layer parameter count by approximately a factor of four.

The widened dense baseline reduces the energy and force MAEs by \(18.1\%\) and \(17.0\%\), 
respectively, relative to the standard 6-layer DPA3 baseline (Fig.~\ref{fig:share_experts}(b)).
As the total number of experts increases, the MoE-E model with two shared experts consistently achieves lower errors than this widened dense baseline.
In particular, with \(N=64\) experts, MoE-E further reduces the energy and force MAEs 
by \(5.8\%\) and \(6.1\%\), respectively, relative to the \(4\times\) Params dense baseline.

These results indicate that, under comparable activated-parameter budgets, 
MoE-E uses parameters more efficiently than uniform dense widening.

\subsection{Element-wise vs Global routing}

We perform a comparative analysis of element-wise routing (MoE-E and MoLE-E) and their global counterparts (MoE-G and MoLE-G) on the OMol25 dataset~\cite{levine2025open}.
This breadth provides a stringent benchmark for evaluating the stability and accuracy of different expert-routing strategies across a highly heterogeneous chemical space.

Table~\ref{tab:GL_results} summarizes the comparative performance of global and element-wise routing architectures across all architectural variants.
In these experiments, we vary the total number of experts $N$ while fixing the number of activated experts to $K=4$, among which two are shared experts.
As shown in Table~\ref{tab:GL_results}, the MoE-G architecture suffers from 
catastrophic training failure: its energy and force MAEs exhibit severe numerical instability and fail to converge.
As a result, its accuracy is not reported.
For the remaining variants (MoE-E, MoLE-E, and MoLE-G), the energy and force MAEs are 
reported after normalization by the corresponding MAEs of the baseline standard DPA3 model.

Overall, increasing the number of experts from $N=4$ to $N=16$ consistently improves both energy and force accuracy across all converged variants.
However, further increasing the number of experts from $N=16$ to $N=64$ yields diminishing returns under the tested settings: performance gains saturate for MoE-E, while accuracy degrades for MoLE-G and for the energy prediction of MoLE-E.
Comparing different mixture designs, MoE-E consistently outperforms both MoLE variants across all expert scales.
Moreover, MoLE-G systematically underperforms its element-wise counterpart MoLE-E, with MoLE-E achieving 
approximately $5\%\text{--}10\%$ lower MAEs in both energy and force predictions across all values of $N$.
Taken together, these results demonstrate that element-wise routing is structurally superior to 
global routing and is essential for the effective and numerically stable integration of MoE architectures into MLIPs.

\subsection{Performance on Multiple Datasets}

To comprehensively evaluate the robustness and generalization capability of the proposed MoE-based interatomic potentials, we benchmark the models on several widely recognized datasets, including OMol25~\cite{levine2025open}, OMat24~\cite{barroso2024open}, and OC20M~\cite{chanussot2021open}. 
These datasets span diverse chemical domains, including molecular systems, extended solids, and catalytic surfaces, enabling a representative evaluation across heterogeneous regimes.

\begin{figure*}
    \centering
    \includegraphics[width=1.0\linewidth]{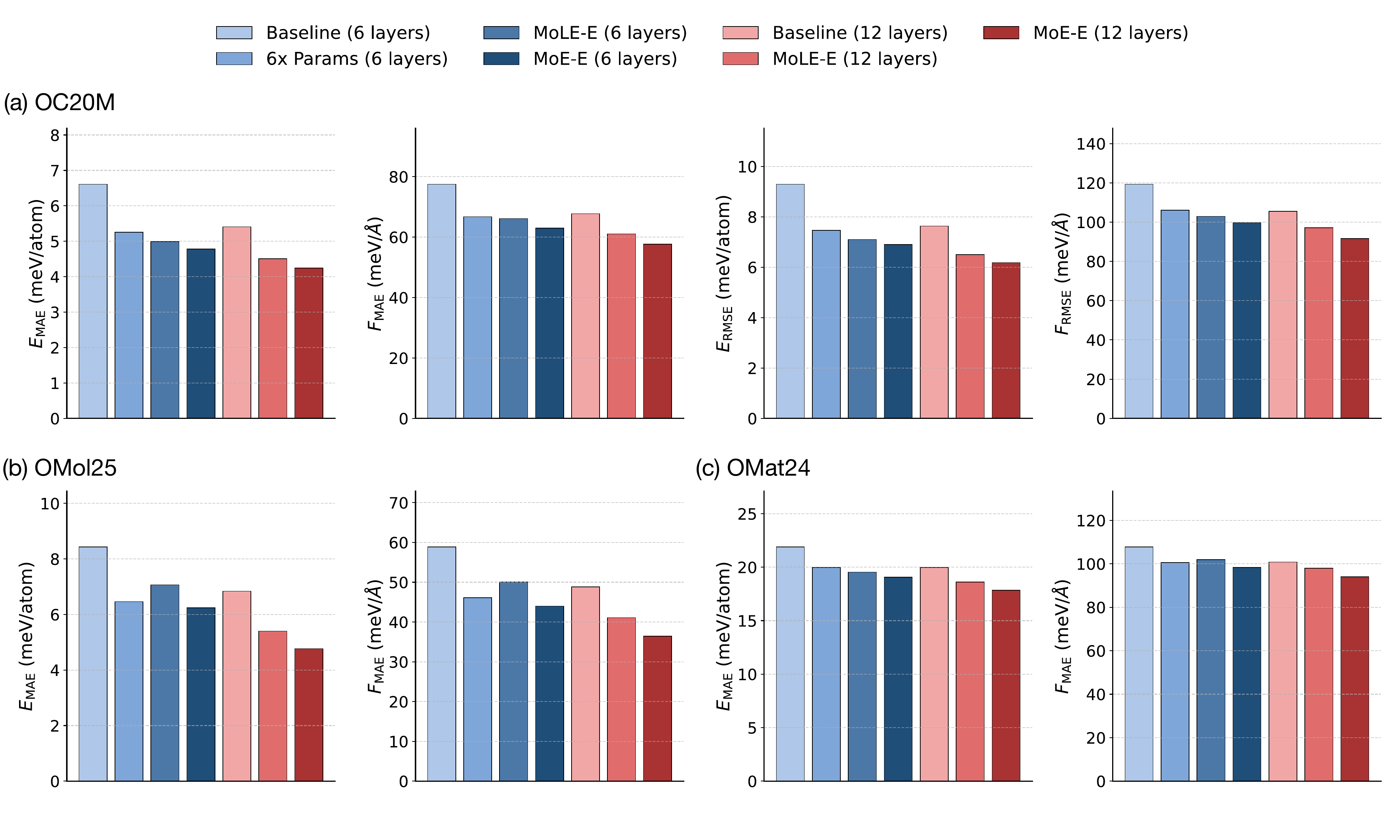}
    \caption{
Performance comparison across three benchmark datasets: (a) OC20M, (b) OMol25, and (c) OMat24. 
For each dataset, energy and force prediction errors are reported in terms of MAE, with additional RMSE shown for OC20M.
We compare baseline DPA3 models with 6-layer and 12-layer configurations against three enhanced variants: a widened dense model (“6$\times$ Params”), MoLE-E, and MoE-E. 
The “6$\times$ Params” model increases the total parameter count by a factor of six through wider hidden layers and is applied only to the 6-layer configuration, enabling comparison with MoE models under a comparable activated parameter budget (with \(K=6\) active experts). 
Both MoLE-E and MoE-E employ 64 experts with 3 shared experts, while MoE-E activates \(K=6\) experts per input. 
The 6-layer baseline, 12-layer baseline, 6$\times$ Params model, MoLE-E, and MoE-E contain approximately 
1.3, 2.6, 7.8, 83, and 83 million parameters, respectively; for MoE-E, approximately 7.8 million 
parameters are activated per forward pass.
}
    \label{fig:multi-dataset}
\end{figure*}

Fig.~\ref{fig:multi-dataset} summarizes the performance across multiple datasets and model scales. 
We primarily report MAE for energy and force predictions across all datasets. 
In addition, for the OC20M benchmark, we also report RMSE to provide further insight into the distribution of prediction errors.

Across all datasets, both 6-layer and 12-layer models exhibit consistent performance improvements when adopting MoE-based architectures compared to their dense counterparts.

In particular, we emphasize the comparison between the MoE models with \(K=6\) activated experts and the widened dense (\(6\times\) Params) baselines, since these configurations have comparable numbers of activated parameters. 
Under this controlled comparison, MoE consistently achieves lower errors than dense enlargement across all datasets, indicating that the sparse expert structure provides more efficient parameter utilization than uniform dense widening.

Comparing different architectures, MoE consistently achieves lower errors than MoLE across all reported metrics. 
This trend holds across both model scales (6-layer and 12-layer), indicating that the advantage of MoE is robust with respect to model depth. 
This performance gap is primarily attributed to the nonlinear formulation in MoE, which provides greater expressivity than the linear structure in MoLE.

Furthermore, Fig.~\ref{fig:multi-dataset} shows that the performance gains exhibit clear dataset-dependent variation. 
On OMol25 and OC20M, MoE achieves substantial relative improvements over the baselines, with energy MAE reduced by approximately 25–30\% and force MAE by about 20–25\%. 
In contrast, the improvements on OMat24 are more moderate, with reductions of around 10–15\% for energy MAE and below 10\% for force MAE.

The observed dataset-dependent variation may reflect an interaction between the DPA3 architecture and the nature of the target systems, rather than being an intrinsic property of the MoE design. 
For instance, the 6$\times$ Params dense baseline also shows substantially smaller gains on OMat24 than on OC20M and OMol25 (Fig.~\ref{fig:multi-dataset}), suggesting that the DPA3 architecture may be less effective at exploiting increased capacity for solid-state materials. 
Nevertheless, MoE consistently outperforms the 6$\times$ Params baseline across all three datasets, consistent with its more efficient use of activated parameters.

For the OC20M dataset, the improvements are observed in both MAE and RMSE metrics, suggesting that MoE not only improves average accuracy but also reduces larger-error cases.

\begin{figure*}
    \centering
    \includegraphics[width=1.0\linewidth]{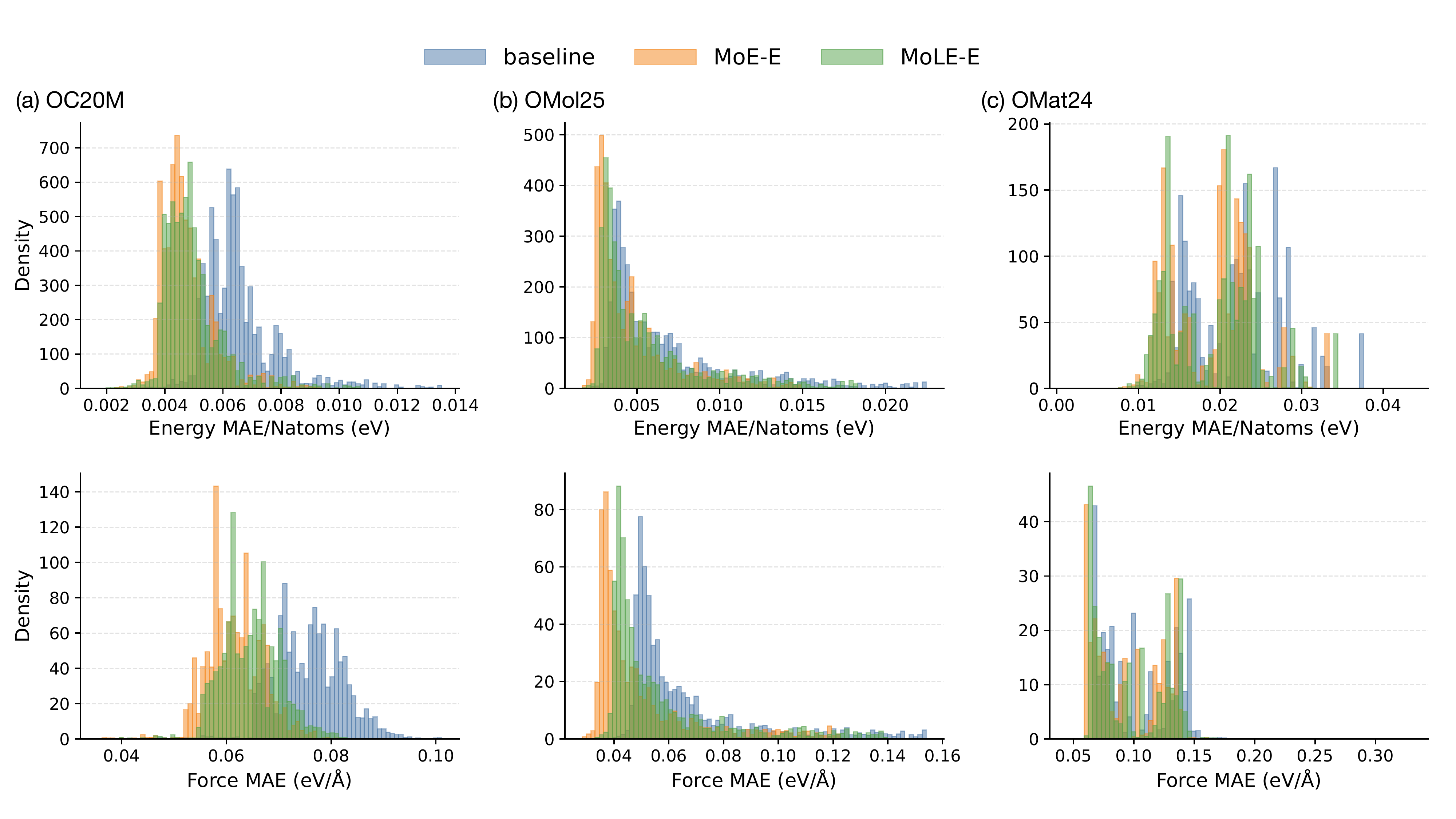}
    \caption{
Distribution of energy and force prediction errors across three benchmark datasets. 
Top row: energy MAE per atom. Bottom row: force MAE. 
Results are shown for (a) OC20M, (b) OMol25, and (c) OMat24. 
Compared to the dense baseline, both MoE-E and MoLE-E shift the error distributions toward lower values and 
significantly suppress high-error tails. MoE-E further produces more concentrated distributions, 
indicating improved stability and more consistent performance across diverse chemical environments. 
All expert-based models use 64 experts with 3 shared experts; for MoE-E, \(K=6\) experts are activated per input. All results correspond to 6-layer models.
}
    \label{fig:MAE_distribution}
\end{figure*}

This observation is further supported by the error distribution analysis shown in Fig.~\ref{fig:MAE_distribution}. 
Across all datasets, the expert-based models (MoE-E and MoLE-E) exhibit a clear shift of the error distributions toward lower values compared to the dense baseline, along with suppressed high-error tails.

On OC20M, the improvement is most pronounced: both the energy and force error distributions exhibit clear leftward shifts together with substantial suppression of the high-error tails, with MoE-E showing the most favorable overall behavior. 
On OMol25, both energy and force distributions are also significantly improved, and MoE-E consistently lies to the left of both the baseline and MoLE-E, indicating better overall accuracy and robustness. 
In contrast, on OMat24, the distributional improvements are more moderate for both energy and force, although expert-based models still shift the distributions toward lower-error regions compared with the baseline.

Notably, MoE-E consistently exhibits a more favorable distribution than MoLE-E, particularly in reducing high error cases, indicating more robust predictions. 
These results confirm that the improvements of MoE-based models extend beyond average metrics and are reflected at the distribution level.

We further verify the reproducibility of the results with respect to random initialization. 
Repeated experiments with different seeds show highly consistent training dynamics and 
negligible variation in final performance, as detailed in Supplementary Note 3.

We further evaluate PES smoothness and MD stability in Supplementary Note~5.
The OMat24 trained MoE-E model produces smooth diatomic energy scans 
without visible discontinuities, supporting that element-wise routing does 
not introduce coordinate-dependent expert switching.
In addition, dataset specific MoE-E checkpoints yield negligible energy drift 
in all tested 50\,ps NVE trajectories, indicating that the sparse expert architecture does not compromise MD stability.

Overall, these results demonstrate that the MoE architecture provides consistent 
improvements across multiple datasets and model scales, while maintaining strong parameter efficiency.

\subsection{Experts weighting Distribution Patterns}

To better understand the behavior of the MoE-E model, we analyze the distribution of routing scores across experts. 
Such analysis provides insights into how the model allocates learning responsibilities and organizes information across different experts.
A natural question underlying this analysis is whether the performance gains of MoE arise from meaningful expert specialization or merely from increased model capacity. 
If the routing scores exhibit structured and interpretable patterns, this would indicate that the routing mechanism reflects systematic specialization rather than arbitrary assignment.

To address this question, we examine the routing scores defined by the projection \(W_e u_i\) before the top-$K$ selection. 
These scores quantify the relative preference of the router over all experts and determine which experts are activated for a given input.

In this analysis, we focus exclusively on routed experts. 
The shared experts, which are uniformly activated across all inputs, are not considered here, as they primarily serve as a global feature backbone and do not reflect input-dependent specialization.

We further investigate whether the learned routing scores correlate with elemental characteristics, which provide a natural organizing principle in interatomic potentials. 
Such correlations would indicate that the model assigns different experts to distinct element types, thereby achieving specialization aligned with chemical identity.

We conduct this analysis on the OMat24 dataset, which provides an ideal testbed due to its extensive coverage of 86 elements with a relatively uniform distribution across the periodic table.

\begin{figure*}
    \centering
    \includegraphics[width=1.0\linewidth]{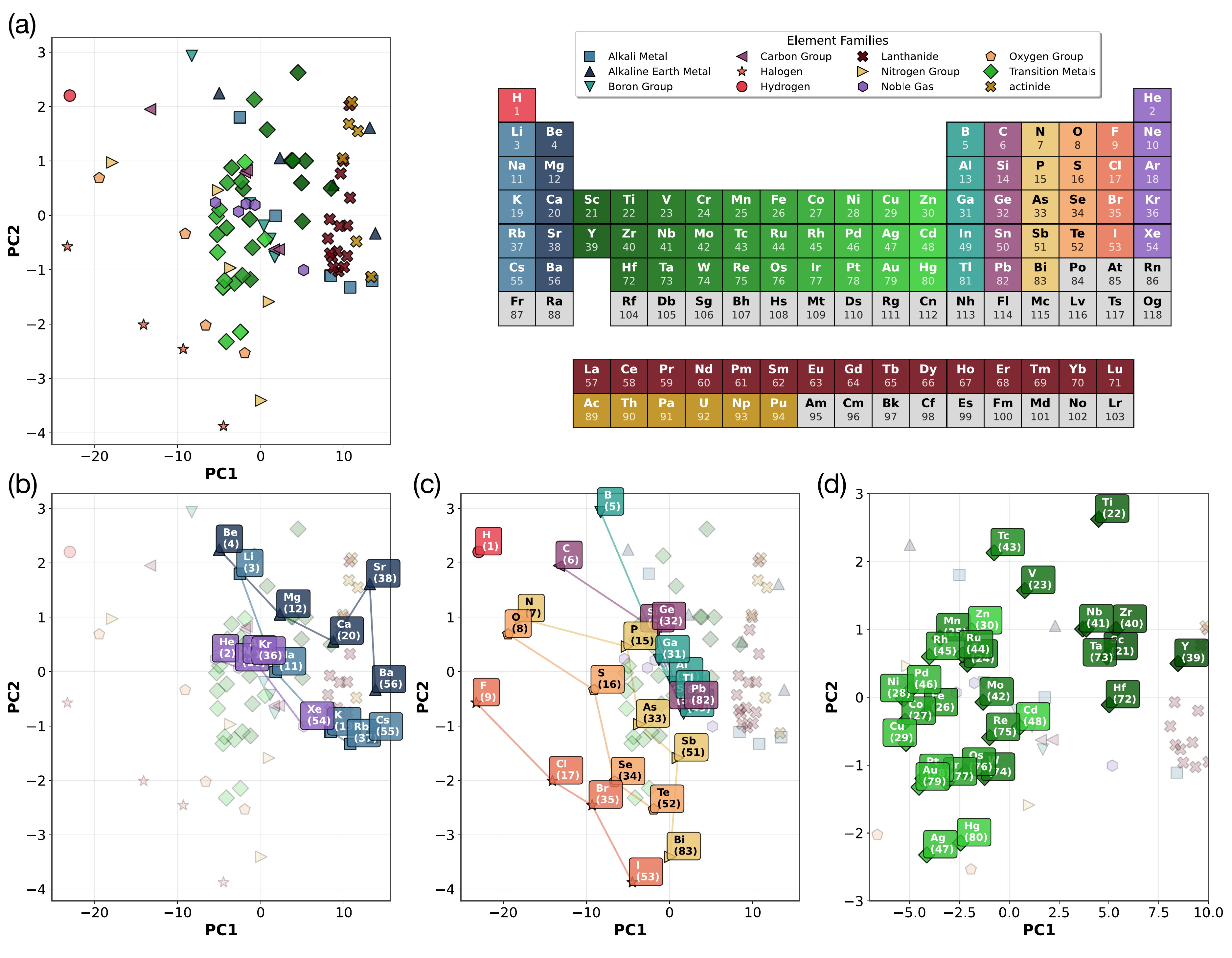}
    \caption{
    Expert weighting distribution analysis via PCA for the MoE-E model (64 total experts, $K=6$ activated experts, 3 shared experts) trained on the OMat24 dataset. (a) Overview of the general distribution for all elements. (b) Distribution patterns for alkali metals, alkaline earth metals, and noble gases. (c) Regional distribution of p-block groups except noble gas, specifically Boron (Group 13) through Halogen (Group 17) elements.(d) Spatial clustering of transition metal elements.
    }
    \label{fig:experts analysis}
\end{figure*}

Principal component analysis (PCA) of the expert-weighting distributions reveals highly efficient dimensionality reduction, with the first two principal components accounting for \(93.57\%\) of the total explained variance. 
Fig.~\ref{fig:experts analysis} presents the projected expert distributions across chemical species, revealing systematic organization in the PCA space~\cite{shazeer2017outrageously, clark2019does}. 
As shown in Fig.~\ref{fig:experts analysis}(a), which illustrates the global distribution of all elements, clear chemical groupings emerge: lanthanides and actinides cluster on the left side of the PCA manifold, while transition metals predominantly occupy the central region.

Fig.~\ref{fig:experts analysis}(b) highlights the alkali metals (Group I), alkaline-earth metals (Group II), and noble gases (Group VIII), while Fig.~\ref{fig:experts analysis}(c) focuses on elements from Groups III to VII together with hydrogen. 
A consistent pattern emerges across chemical groups: elements within the same group arrange themselves diagonally from the top-left to the bottom-right of the PCA space. 
This spatial ordering correlates strongly with increasing atomic number, with lighter elements located in the top-left region and heavier elements positioned toward the bottom-right.

Along the period direction, as shown in Fig.~\ref{fig:experts analysis}(c), elements with pronounced non-metallic character exhibit a systematic shift from the top-right toward the bottom-left of the PCA space with increasing atomic number.
In contrast, strongly metallic elements follow the opposite orientation, extending from the bottom-left (Group I) to the top-right (Group II), as visualized in Fig.~\ref{fig:experts analysis}(b). 
Elements with intermediate chemical behavior—particularly those in Groups III (Al, Ga, In, Tl) and IV (Si, Ge, Sn, Pb)—display overlapping distributions in the PCA embedding (Fig.~\ref{fig:experts analysis}(c)). 
Noble gases, although electronically distinct, constitute an exception: their embeddings partially overlap with those of intermediate elements, which may reflect their relatively limited representation in the OMat24 dataset.

As illustrated in Fig.~\ref{fig:experts analysis}(d), transition metals are predominantly concentrated near the center of the PCA space. 
Within this cluster, elements belonging to the same period exhibit a clear diagonal alignment: as the group number increases, their distributions shift from the top-right toward the bottom-left. 
This trend is further reflected by the color gradient from dark to light green, capturing the systematic evolution of their electronic characteristics.

These structured distributions indicate that the MoE-E router encodes elemental characteristics in a manner consistent with known physicochemical trends. 
The observed organization—such as group-wise alignments, periodic progressions, and the clustering of transition metals and \(f\)-block elements—suggests that the model learns a chemically meaningful representation of the periodic table.

This behavior provides evidence of element-wise expert specialization, where different experts are preferentially activated for distinct element types. 
Such specialization offers a plausible explanation for the improved accuracy and generalization observed in MoE-E.

By leveraging elemental identity in the routing mechanism, the model is able to better distinguish among chemically diverse species while allocating expert capacity more effectively across compositional space. 
Overall, this result highlights the potential of MoE-based designs to combine learned representations with chemical intuition, improving both performance and interpretability in MLIPs.

\subsection{Transferability to CHGNET}\label{subsec:chgnet}

To assess whether the MoE design transfers to MLIP architectures beyond DPA3, 
we applied the element-wise MoE-E framework to CHGNET~\cite{deng2023chgnet}, 
a graph neural network potential with a different network topology and featurization 
from DPA3. Because the full CHGNET dimensions led to excessive memory consumption 
when combined with the MoE expert pool, the feature and hidden dimensions were halved 
from the original setting; the comparison is therefore between a matched halved-dimension 
CHGNET baseline and its MoE-E extensions. We use a pool of $N=32$ total experts and vary the 
number of activated experts $K=2,4,6$, with shared experts set to $1,2,3$, respectively. 
The models were trained on the MPtrj dataset for 20 epochs using energy, force, and stress labels.

As reported in Supplementary Table~7, the MoE-E extensions improve energy, 
force, and stress predictions on the MPtrj test set relative to the matched 
CHGNET baseline, and the gains increase monotonically with $K$. For $K=6$, 
the test energy MAE is reduced from $35.9$ to $32.4$ meV/atom ($\sim$9.7\%), 
the test force MAE from $80.8$ to $75.9$ meV/\AA\ ($\sim$6.1\%), and the test 
stress MAE from $0.3636$ to $0.3468$ GPa ($\sim$4.6\%). 
These results indicate that the element-wise MoE strategy 
is not restricted to DPA3, while the specific design rules 
identified in this work may still depend on the backbone size, 
feature representation, and dataset. A more systematic study of 
full-scale CHGNET and equivariant MLIP backbones is left for future work.

\section{Discussion}

In this work, we present a systematic investigation of MoE architectures for MLIPs and demonstrate that sparsely 
activated expert enlargement provides an effective pathway for improving predictive accuracy beyond conventional dense model enlargement.
By integrating MoE designs into the DPA3 framework, we demonstrate that sparse activation, shared-expert mechanisms, nonlinear expert specialization, 
and element-wise expert routing collectively contribute to consistent reductions in energy and force prediction errors.
Across diverse benchmarks—including OMol25, OMat24, and OC20M—the proposed element-wise MoE architecture consistently achieves 
substantial improvements in both energy and force prediction accuracy, while also exhibiting interpretable expert specialization patterns that align with periodic chemical trends.

Our analysis provides several conceptual insights into MoE design choices for developing atomistic foundation models. 
First, the introduction of sparse activation together with shared experts enables the learning of transferable 
representations of common interatomic interactions, allowing efficient utilization of large expert pools. 
Second, nonlinear expert aggregation delivers greater representational capacity than linear mixture formulations, 
underscoring the importance of nonlinear specialization for accurately modeling complex potential-energy surfaces. 
Third, element-wise routing is essential for avoiding training instability and enabling effective specialization in chemically heterogeneous systems, 
consistently outperforming global routing strategies that may introduce optimization instability. 
Collectively, these findings establish MoE-based conditional computation as a principled alternative 
to dense parameter enlargement for next-generation large atomistic models.

Despite these advances, an important limitation of the present study is that the current implementation 
does not yet exploit a fully distributed training and inference framework specifically optimized for sparse expert parallelism. 
While MoE architectures are intrinsically well-suited for distributed environments and offer substantial theoretical 
scalability advantages over dense scaling, realizing these benefits in practice requires dedicated expert-parallel 
system design, communication scheduling, and load-balancing strategies. Future work will therefore focus on developing 
large-scale distributed MoE training and inference pipelines to unlock the full 
computational efficiency and scaling potential of expert-based atomistic foundation models.

Overall, this work demonstrates that MoE architectures provide a promising and physically 
grounded paradigm for expanding the expressive capacity of MLIP, offering a foundation for 
increasingly accurate and transferable simulations across the chemical and materials sciences.

\section{Methods}\label{sec: method}

The architectural details and training protocols used in the above experiments are described below.

\subsection{Detailed architecture formulations}
The MoE and MoLE transformations introduced in Results replace the dense layer of Eq.~\eqref{eq:1} by the formulations detailed below.

In this work, we use the DPA3 model architecture to illustrate the incorporation of a MoE design into a MLIP. 
The same approach can be readily extended to other graph neural networks that employ invariant features, without substantial modification. 
The model takes as input the atomic coordinates and chemical species, and outputs the total potential energy \(E\) of the system.
DPA3 assumes an atom-wise energy decomposition,
\begin{equation}
    E = \sum_i E_i, \quad E_i = E_i\!\left(v_i^{1, L}\right),
\end{equation}
where $E_i$ denotes the energy contribution of atom $i$, which is derived from the final-layer vertex feature $v_i^{1,L}$. 
This feature represents the state of atom $i$ within the primary graph of the line-graph series after $L$ update layers.
The majority of DPA3 model parameters reside in the dense layers used to construct messages during message passing along graph edges, which connect atom pairs whose interatomic distance falls within a predefined cutoff radius.
A detailed description of the DPA3 architecture is provided in Ref.~\cite{zhang2025graph}. 

For each input feature \(x_i\) associated with atom \(i\), a dense layer computes
\begin{equation}
    y_i = \sigma\!\left(W x_i + b\right),
    \label{eq:1}
\end{equation}
where \(W\) and \(b\) denote the weight matrix and bias vector, respectively, \(\sigma\) is a nonlinear activation function, and \(y_i\) is the resulting output (see also Fig.~\ref{fig:MoE}(a)). 
The representational capacity of the model can be increased by enlarging these dense layers, i.e., by increasing the dimensionality of the weight and bias parameters.

Assume that the MoE contains a total of \(N\) experts, consisting of \(I\) routed experts and \(N-I\) shared experts.
The MoE transformation is defined as
\begin{align}
   y_i &= \sum_{j=1}^{I} \alpha_{ij}\,
   \sigma\!\left(W_{j} x_i + b_j\right)
   + \sum_{j=I+1}^{N}
   \sigma\!\left(W_{j} x_i + b_j\right),
   \label{eq:MoE-E:1} \\
   \alpha_{ij} &= 
   \begin{cases}
   s_{ij}, & \text{if } s_{ij} \in \mathrm{TopK}\!\left(
   \{ s_{im}\},\, K^{\prime} \right), \\
   0, & \text{otherwise},
   \end{cases}
   \label{eq:MoE-E:2} \\
   s_{ij} &= 
   \mathrm{softmax}
   \!\left(W_e u_i\right),
   \label{eq:MoE-E:3}
\end{align}
where \(\sigma(\cdot)\) denotes a nonlinear activation function, \(W_j\) and \(b_j\) are the trainable expert-specific weights and bias, and \(\alpha_{ij}\) is the gating weight applied to the \(j\)-th routed expert for atom \(i\).
The operator \(\mathrm{TopK}\!\left(\{ s_{im} \}, K^{\prime}\right)\) selects the largest \(K^{\prime} \le I\) elements from the set of gating scores \(\{ s_{im} \mid 1 \le m \le I \}\), thereby enforcing sparse expert activation.
Including the $(N-I)$ shared experts that are always active, the total number of activated experts is denoted by $K$, satisfying
\begin{equation}
K = K^{\prime} + (N - I).
\end{equation}
The gating scores \(s_{ij}\) are computed according to \eqref{eq:MoE-E:3}, where \(W_e \in \mathbb{R}^{I \times M}\) is a trainable routing matrix and \(u_i \in \mathbb{R}^{M}\) is a latent representation of dimension $M$, encoding the chemical identity of atom \(i\).

The hidden representation \(u_i\) is defined as
\begin{equation}
u_i = \mathrm{MLP}\!\left(\mathrm{one\_hot}(Z_i)\right),
\label{eq:element-wise router}
\end{equation}
where the atomic number \(Z_i\) is first mapped to a one-hot vector and subsequently projected into a continuous embedding space via a multilayer perceptron (MLP).
This element-wise routing mechanism enables atom-type–dependent expert selection while preserving smoothness and differentiability, which are essential for stable and physically consistent interatomic potential modeling.

In the MoLE formulation, the features are first linearly transformed by expert-specific weights, after which the expert contributions are aggregated via gating weights, and a nonlinear activation function is applied at the end.
This yields
\begin{align}
\label{eq:MoLE-E:1}
&y_i = \sigma\!\left(
\sum_{j=1}^{I} \alpha_{ij} \left( W_j x_i + b_j \right)
+ \sum_{j=I+1}^{N} \left( W_j x_i + b_j \right)
\right), \\ 
\label{eq:MoLE-E:2}
   &\alpha_{ij} = 
   \mathrm{softmax}
   \bigg(W_e u_i \bigg) 
\end{align}
where \(\sigma(\cdot)\) denotes the activation function.
Equation~\ref{eq:MoLE-E:1} is mathematically equivalent to a single dense layer with effective parameters,
\begin{align}
y_i &= \sigma\!\left( \bar W x_i + \bar b \right), \\
\bar W &= \sum_{j=1}^{I} \alpha_{ij} W_j + \sum_{j=I+1}^{N} W_j, \\
\bar b &= \sum_{j=1}^{I} \alpha_{ij} b_j + \sum_{j=I+1}^{N} b_j.
\end{align}
Thus, MoLE can be interpreted as a data-dependent linear transformation whose weights and biases are modulated by the gating mechanism prior to the nonlinearity.

\subsection{Training Protocol}
We evaluate our models on three benchmark datasets: OMol25~\cite{levine2025open}, OMat24~\cite{barroso2024open}, and OC20M~\cite{chanussot2021open}. 
For OMol25 and OMat24, we adopt the training and validation splits provided in the original publications. 
For OC20M, we use the training and validation splits released by AIS Square\cite{aissquare_openlam}

Detailed training hyperparameters are provided in Supplementary Table 1. 
For reference, the standard 6-layer DPA3 baseline contains approximately 1.3 million parameters, 
predominantly in the dense message-passing layers. The 12-layer baseline contains approximately 
2.6 million parameters, and the 4$\times$ and 6$\times$ widened dense baselines contain approximately 
5.2 and 7.8 million parameters, respectively. In the MoE and MoLE models, each expert replaces a 
dense layer block of comparable capacity; consequently, a model with $N$ experts contains approximately 
$1.3N$ million total parameters, and activating $K$ experts corresponds to approximately $1.3K$ million 
activated parameters per forward pass. For example, the MoE-E model with 64 total experts and $K=6$ 
activated experts (3 shared) has approximately 83 million total parameters and approximately 7.8 million 
activated parameters. 

To ensure a controlled comparison, all models on the same dataset are trained for a fixed budget of 1 million steps. 
Across different experiments, we vary the MoE-related 
architectural components as well as the backbone width and depth when constructing models of different scales.

All experiments are conducted using single-precision training on NVIDIA H20 and A800 GPUs.

\backmatter

\bmhead{Data Availability}
\sloppy
% All data supporting the findings of this study are available within the Supplementary Information
The datasets used in this work, including OMat24, OC20M, and OMol25, 
are publicly available and can be accessed from their original 
publications~\cite{barroso2024open, chanussot2021open, levine2025open} and 
AIS Square (\url{https://www.aissquare.com/datasets/detail?pageType=datasets&name=OpenLAM-TrainingSet-v1&id=308}).

\bmhead{Code Availability}
The implementation of the MoE model used in this work is available at \url{https://github.com/iProzd/deepmd-kit/tree/moe}.

\bmhead{Acknowledgements}
Computational resources utilized in this work were provided by the AI for Science Institute. 
The work  is supported by the National Key R\&D Program of China (Grant No.~2022YFA1004300) 
and the National Natural Science Foundation of China (Grant Nos.~12525113 and~12561160120).

\bmhead{Author Contribution}
Y.L., D.Z., A.P., and H.W. conceived the idea of this work, designed the model structure, 
and implemented the model. The experiments were mainly designed and performed by Y.L. D.Z. and A.P. 
performed data collection and model tests on different systems. W.E., L.Z., and H.W. supervised the project. 
All authors contributed to the discussions and edited the manuscript. 
All authors have read and approved the final manuscript.

\bmhead{Competing Interests}
The authors declare no competing financial or non-financial interests.

\bibliography{sn-bibliography}

%=================================================%
% Figures and Tables moved to end of manuscript %
%=================================================%

\begin{table*}[]
\centering
\begin{tabular}{l c ccc c ccc}
\toprule
\multirow{2}{*}{\textbf{Model}} & \multirow{2}{*}{\textbf{Routing}} & \multicolumn{3}{c}{\textbf{Normalized Energy MAE}} & & \multicolumn{3}{c}{\textbf{Normalized Force MAE}} \\
\cmidrule{3-5} \cmidrule{7-9}
& & 4 Experts & 16 Experts & 64 Experts & & 4 Experts & 16 Experts & 64 Experts \\
\midrule
\multirow{2}{*}{MoE}  
& Element-wise (E) & 0.814 & 0.780 & 0.771 & & 0.809 & 0.793 & 0.779 \\
& Global (G) & N/A & N/A & N/A & & N/A & N/A & N/A \\
                    
\midrule
\multirow{2}{*}{MoLE} 
& Element-wise (E) & 0.913 & 0.847 & 0.866 & & 0.914 & 0.857 & 0.847 \\
& Global (G) & 0.981 & 0.946 & 0.952 & & 0.965 & 0.936 & 0.950 \\
                    
\bottomrule
\end{tabular}
\caption{Normalized Energy and Force MAE values. The results compare the performance of Global (G) vs. 
Element-wise (E) routing across different expert counts. All MAEs are normalized against a 
6-layer DPA3 baseline; 
thus, values represent relative error ratios where a lower score signifies superior performance relative to the baseline. 
}
\label{tab:GL_results}
\end{table*}

\end{document}

\tableofcontents
\section{Hyperparameters and System Configurations}
The baseline DPA3 model utilizes hyperparameters summarized in Table~\ref{tab:hyperparams}, 
following the standard configuration established in the DPA3 paper \cite{zhang2025graph}. 
However, batch sizes were adjusted according to the dataset: a batch size of 256 was used 
for OMol25, while 128 was used for both OMat24 and OC20M.
\begin{table*}[htbp]
  \centering
  \begin{threeparttable}
  \caption{Hyperparameters for 6-layer and 12-layer DPA3 model.}
  \label{tab:hyperparams}
  \setlength{\tabcolsep}{20pt}  
  \renewcommand{\arraystretch}{1.2} 
  \setlength{\arrayrulewidth}{0.6pt}  
  \begin{tabular}{l|c}
    \toprule
    \textbf{Hyperparameters} &
    \textbf{baseline} \\
    \hline
    Number of update layers $L$ & 12/6\\
    Maximum graph order $k$ & 2 \\
    Dimension of vertex features in $G^{(0)}$ & 128 \\
    Dimension of vertex features in $G^{(1)}$ & 64 \\
    Dimension of vertex features in $G^{(2)}$ & 32 \\
    Cutoff radius in $G^{(1)}$ $\AA$ & 6.0 \\
    Cutoff radius in $G^{(2)}$ $\AA$ & 4.0 \\
    Batch size & 128/256 \\
    Optimizer & AdamW  \\
    Learning rate scheduling & Exp \\
    Maximum learning rate & 1e-3 \\
    Minimum learning rate & 1e-5 \\
    Number of training steps & 1M \\
    Activation Function Threshold & 10.0 \\
    Loss function & MSE \\
    \bottomrule
  \end{tabular}
  \end{threeparttable}
\end{table*}

The model dimensions are scaled to create larger baselines: for the ``\(4\times\) params baseline'', 
the dimensions of $G^{(0)}$, $G^{(1)}$, and $G^{(2)}$ are doubled from 128, 
64, and 32 to 256, 128, and 64, respectively. Similarly, the ``\(6\times\) params baseline'' 
is constructed by expanding these dimensions to 314, 156, and 78.

The hyperparameters for the MoE and MoLE models are identical to the baseline model. 
Further implementation details specific to the MoE and MoLE architectures are provided in the main text.

\section{Energy and Force MAE value}
While the architectural ablation results in the main text are reported as normalized MAEs, 
this section provides the absolute MAE values for direct comparison. 
Specifically, Table~\ref{tab:MoE performance} details the energy and force 
MAEs for the MoE-E model across different numbers of activated experts $K$, corresponding to the trends visualized in Fig. 2(a).
\begin{table*}[htbp]
  \centering
  \caption{Performance comparison of different configurations with varying numbers of shared experts.}
  \label{tab:MoE performance}
  \setlength{\tabcolsep}{12pt}
  \renewcommand{\arraystretch}{1.2}
  \begin{tabular}{lccc}
    \toprule
    \textbf{MoE-E} &
    \textbf{Shared Experts} &
    \textbf{Energy MAE} (\unit{\electronvolt}) &
    \textbf{Force MAE} (\unit{\electronvolt\per\angstrom}) \\
    \midrule
    Baseline & -- & 8.43e-3 & 5.89e-2 \\
    \midrule
    
    \multirow{2}{*}{$K=2$}
      & 0 & 9.12e-3 & 5.68e-2 \\
      & 1 & 7.92e-3 & 5.37e-2 \\
    
    \midrule
    \multirow{3}{*}{$K=4$}
      & 0 & 8.14e-3 & 5.41e-2 \\
      & 2 & 6.50e-3 & 4.59e-2 \\
      & 3 & 6.47e-3 & 4.65e-2 \\
    
    \midrule
    \multirow{4}{*}{$K=6$}
      & 0 & 7.78e-3 & 5.31e-2 \\
      & 2 & 5.99e-3 & 4.51e-2 \\
      & 3 & 6.24e-3 & 4.40e-2 \\
      & 4 & 6.16e-3 & 4.41e-2 \\
    
    \midrule
    \multirow{4}{*}{$K=8$}
      & 0 & 7.81e-3 & 5.30e-2 \\
      & 2 & 6.15e-3 & 4.47e-2 \\
      & 4 & 5.68e-3 & 4.21e-2 \\
      & 6 & 5.54e-3 & 4.13e-2 \\
    \bottomrule
  \end{tabular}
\end{table*}

Similarly, Table~\ref{tab:MoLE performance} summarizes the performance metrics for both MoLE-E and MoE-E. This table provides the numerical data supporting the trends visualized in Fig. 2(b) of the main text.
Table~\ref{tab:GL vs Element} compares Global and Element-wise routing, offering the specific values corresponding to the summary in Table 1.
\begin{table*}[htbp]
  \centering
  \caption{Performance comparison of different configurations with varying numbers of shared experts.}
  \label{tab:MoLE performance}
  \setlength{\tabcolsep}{12pt}
  \renewcommand{\arraystretch}{1.2}
  \begin{tabular}{c|cccc}
    \toprule
    \textbf{Structure} &
    \textbf{Share Experts} &
    \textbf{Experts} &
    \textbf{Energy MAE} (\unit{\electronvolt}) &
    \textbf{Force MAE} (\unit{\electronvolt\per\angstrom}) \\
    \hline
    Baseline & -- & -- & 8.43e-3 & 5.89e-2 \\
    \hline
    4$\times$ params Baseline & -- & -- & 6.90e-3 & 4.89e-2 \\
    \hline
    
    \multirow{6}{*}{MoLE-E} & \multirow{3}{*}{0}
      & 4 & 8.65e-3 & 5.74e-2 \\
      & & 16 & 8.10e-3 & 5.52e-2 \\
      & & 64 & 8.29e-3 & 5.50e-2 \\
    % \addlines[2pt]
      \cline{2-5}  
      & \multirow{3}{*}{2}
      & 4 & 7.70e-3 & 5.38e-2 \\
      & & 16 & 7.14e-3 & 5.04e-2 \\
      & & 64 & 7.30e-3 & 4.99e-2 \\
    \hline
    \multirow{6}{*}{$K=4$ MoE-E} & \multirow{3}{*}{0}
      & 4 & 8.80e-3 & 5.59e-2 \\
      & & 16 & 8.12e-3 & 5.43e-2 \\
      & & 64 & 8.14e-3 & 5.41e-2 \\
    % \addlines[2pt]
      \cline{2-5}  
      & \multirow{3}{*}{2}
      & 4 & 6.86e-3 & 4.76e-2 \\
      & & 16 & 6.58e-3 & 4.67e-2 \\
      & & 64 & 6.50e-3 & 4.59e-2 \\
    \bottomrule
  \end{tabular}
\end{table*}

\begin{table*}[htbp]
\centering
\caption{Energy and Force MAE values. The results compare the performance of Global (G) vs. Element-wise (E) routing across different expert counts.}
\begin{tabular}{l c ccc c ccc}
\toprule
\multirow{2}{*}{\textbf{Model}} & \multirow{2}{*}{\textbf{Routing}} & \multicolumn{3}{c}{\textbf{Energy MAE}(eV)} & & \multicolumn{3}{c}{\textbf{Force MAE} (eV $\AA^{-1}$)}\\
\cmidrule{3-5} \cmidrule{7-9}
& & 4 Experts & 16 Experts & 64 Experts & & 4 Experts & 16 Experts & 64 Experts \\
\midrule
\multirow{2}{*}{MoLE} & Global (G) & 8.27e-3 & 7.98e-3 & 8.03e-3 & & 5.68e-2 & 5.51e-2 & 5.59e-2 \\
                      & Element-wise (E) & 7.70e-3 & 7.14e-3 & 7.30e-3 & & 5.38e-2 & 5.04e-2 & 4.99e-2 \\
\midrule
\multirow{2}{*}{MoE}  & Global (G) & N/A & N/A & N/A & & N/A & N/A & N/A \\
                      & Element-wise (E) & 6.86e-3 & 6.58e-3 & 6.50e-3 & & 4.76e-2 & 4.67e-2 & 4.59e-2 \\
\bottomrule
\end{tabular}
\label{tab:GL vs Element}
\end{table*}

\begin{figure}
    \centering
    \includegraphics[width=1.0\linewidth]{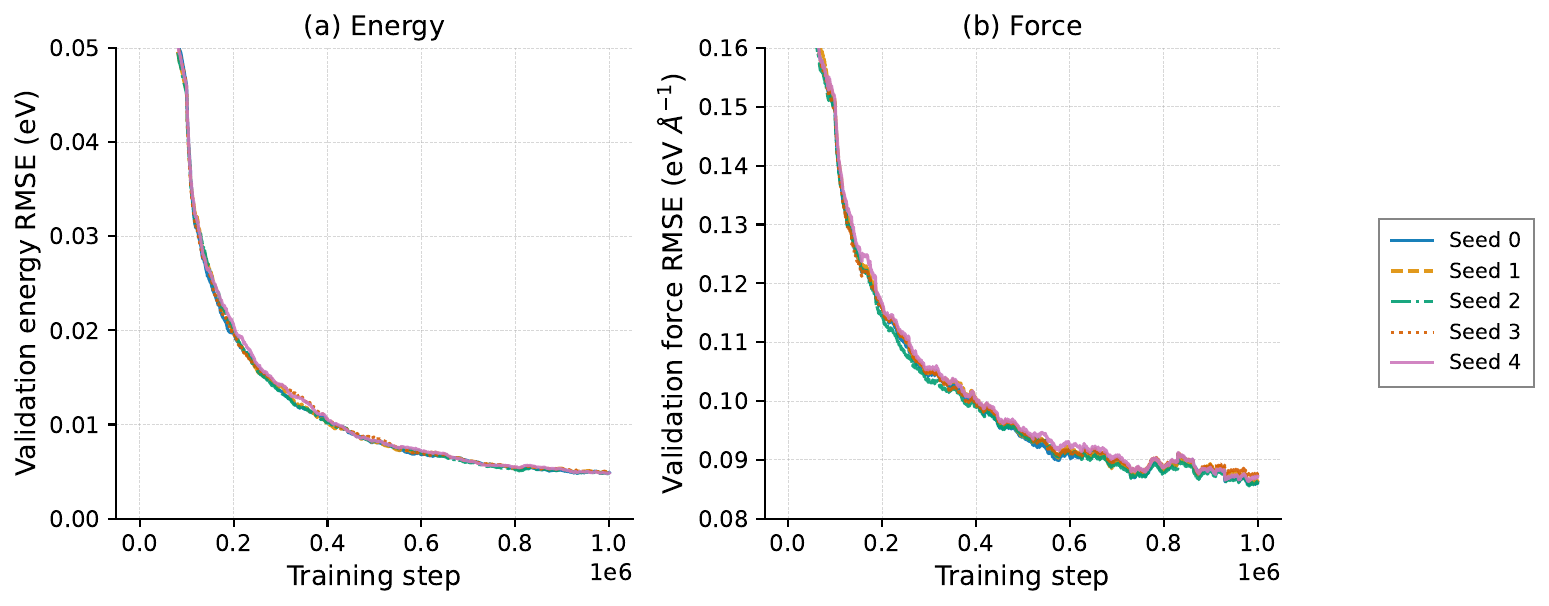}
    \caption{
Training curves under different random seeds on the OC20M dataset using the MoE-E model (64 experts, 3 shared experts, $K=6$). 
(a) Validation energy RMSE and (b) validation force RMSE as functions of training steps. 
The curves from different seeds show nearly identical convergence behavior, demonstrating strong training stability and reproducibility.
}
    \label{fig:seed_variation}
\end{figure}
\section{Reproducibility Across Random Seeds}
To further assess the robustness of the training process, 
we perform repeated experiments with different random seeds on the OC20M dataset using the MoE-E model with 64 experts, 3 shared experts, and $K=6$ activated experts.

As shown in Supplementary Fig.~\ref{fig:seed_variation}, 
both the validation energy RMSE and force RMSE exhibit highly consistent convergence behavior across all runs. 
The training curves corresponding to different seeds almost 
overlap throughout the entire optimization process, indicating that the training dynamics are largely insensitive to random initialization and data shuffling.

Moreover, the variance in the final validation errors across 
different seeds is negligible compared to the performance differences between model architectures reported in the main text. 
This demonstrates that the observed improvements of the 
MoE-based models are stable and reproducible, rather than arising from stochastic variations in training.

\section{Training Convergence}
To ensure that the architectural comparisons reported in the main 
text are not biased by under convergence, we monitor the validation 
force MAE of all models at 900k and 1000k training steps. 
The relative change in force MAE over the final 100k steps is defined as
\begin{equation}
    \Delta F_{\mathrm{MAE}} = \frac{F_{\mathrm{MAE}}^{1000k} - F_{\mathrm{MAE}}^{900k}}{F_{\mathrm{MAE}}^{900k}} \times 100\% .
\end{equation}
As summarized in Supplementary Table~\ref{tab:convergence}, 
the force MAE changes by less than 1.3\% over 
the final 100k steps for every model, with most models showing relative changes below 1\%. 
This indicates that the force predictions have largely converged by 1M steps.

The OMol25 MoE-E model with $K=8$ shows the largest remaining force improvement (1.22\%), 
suggesting that it is approaching but has not fully reached convergence; 
this is expected because a larger number of activated experts increases the 
effective model capacity per forward pass and can slow late stage convergence. 
Overall, the results support the use of 1M training steps as a consistent and 
adequate convergence criterion across the ablation studies.

\begin{table*}[htbp]
  \centering
  \caption{Validation force MAE at 900k and 1000k training steps. The MoE-E models use 64 total experts.}
  \label{tab:convergence}
  \setlength{\tabcolsep}{5pt}
  \renewcommand{\arraystretch}{1.2}
  \small
  \begin{tabular}{llccc}
    \toprule
    \textbf{Dataset} & \textbf{Model} & \textbf{Step} & \textbf{$F_{\mathrm{MAE}}$ (eV/\AA)} & \textbf{$\Delta F_{\mathrm{MAE}}$ (\%)} \\
    \midrule
    \multirow{4}{*}{OC20M}
    & \multirow{2}{*}{MoE-E ($K$=6, 3-shared)} & 900k & 0.0632 & \\
    & & 1000k & 0.0630 & -0.41 \\
    \cmidrule{2-5}
    & \multirow{2}{*}{Baseline} & 900k & 0.0777 & \\
    & & 1000k & 0.0775 & -0.23 \\
    \midrule
    \multirow{4}{*}{OMat24}
    & \multirow{2}{*}{MoE-E ($K$=6, 3-shared)} & 900k & 0.0987 & \\
    & & 1000k & 0.0983 & -0.32 \\
    \cmidrule{2-5}
    & \multirow{2}{*}{Baseline} & 900k & 0.1080 & \\
    & & 1000k & 0.1078 & -0.20 \\
    \midrule
    \multirow{10}{*}{OMol25}
    & \multirow{2}{*}{MoE-E ($K$=2, 1-shared)} & 900k & 0.0542 & \\
    & & 1000k & 0.0537 & -0.82 \\
    \cmidrule{2-5}
    & \multirow{2}{*}{MoE-E ($K$=4, 2-shared)} & 900k & 0.0461 & \\
    & & 1000k & 0.0459 & -0.51 \\
    \cmidrule{2-5}
    & \multirow{2}{*}{MoE-E ($K$=6, 3-shared)} & 900k & 0.0443 & \\
    & & 1000k & 0.0440 & -0.68 \\
    \cmidrule{2-5}
    & \multirow{2}{*}{MoE-E ($K$=8, 4-shared)} & 900k & 0.0426 & \\
    & & 1000k & 0.0421 & -1.22 \\
    \cmidrule{2-5}
    & \multirow{2}{*}{Baseline} & 900k & 0.0590 & \\
    & & 1000k & 0.0589 & -0.19 \\
    \bottomrule
  \end{tabular}
\end{table*}

\section{PES smoothness and MD stability}

To assess whether element-wise sparse routing introduces non-smooth 
behavior in the predicted potential-energy surface or compromises molecular dynamics stability, we performed two complementary tests: 
diatomic PES scans and NVE molecular dynamics simulations.

\begin{figure}
    \centering
    \includegraphics[width=1.0\linewidth]{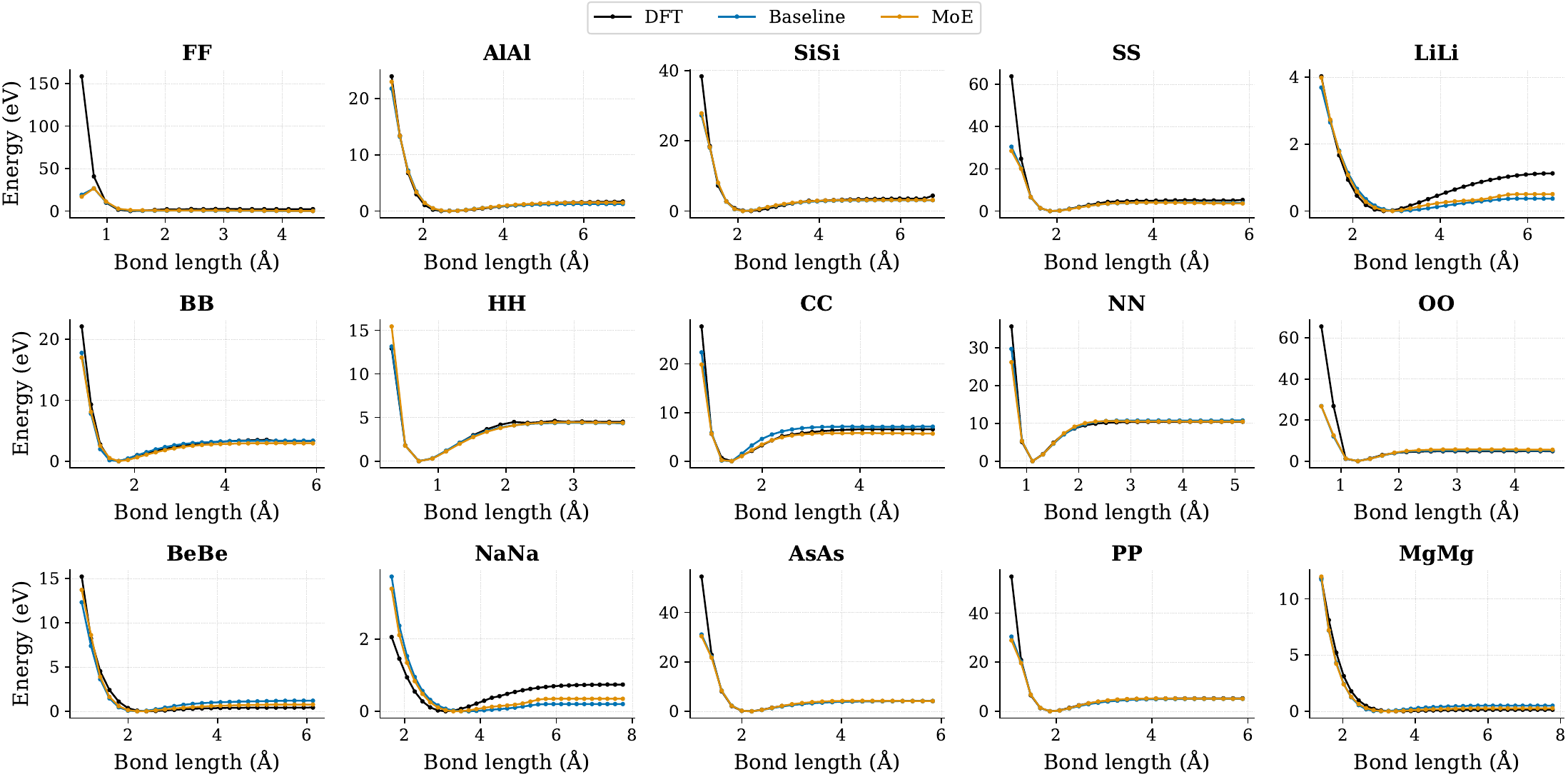}
    \caption{
  Diatomic potential-energy 
  curves for fifteen homonuclear dimers 
  as a function of bond length, computed with DFT (black), 
  the dense DPA3 baseline (blue), and the MoE-E model (orange).
  The MoE-E model was trained on OMat24 with 64 total experts, 
  $K=6$ activated experts, and 3 shared experts.
  The curves are smooth and continuous across the scanned range 
  for all three methods; near equilibrium the MoE-E and baseline
   predictions closely track the DFT reference, while at stretched 
   bond lengths both show comparable deviations for some dimers.
  }
    \label{fig:pes_smoothness}
\end{figure}

For the PES smoothness test, we used the MoE-E checkpoint trained on OMat24 with 64 total experts,
\(K=6\) activated experts, and 3 shared experts. We computed diatomic potential-energy curves by continuously
varying the interatomic distance for fifteen representative homonuclear dimers.
Supplementary Fig.~\ref{fig:pes_smoothness} compares the MoE-E curves with those from the dense DPA3 baseline and DFT.
All three methods produce smooth, continuous curves across the scanned range.
Near the equilibrium distance the MoE-E and baseline predictions closely track the 
DFT reference; at stretched bond lengths both the MoE-E and baseline curves deviate from DFT for some dimers, with no abrupt jumps 
or discontinuities observed in the MoE-E trajectories.

For the MD stability test, we performed 50\,ps NVE simulations using both the dense 
DPA3 baseline and the corresponding MoE-E model. For each dataset, the models were trained on the same dataset as the
evaluated validation structures. Specifically, OMol25, OMat24, and 
OC20M trajectories were evaluated using the corresponding OMol25-trained, OMat24-trained, and OC20M-trained checkpoints. In all cases, the MoE-E 
model used 64 total experts, \(K=6\) activated experts, and 
3 shared experts. For each dataset, three structures were randomly selected from the validation set, giving nine trajectories in total.

Following the LAMBench protocol~\citep{peng2026lambench}, 
MD stability was quantified by the absolute value of the energy drift slope. For each NVE trajectory, the total energy was recorded throughout the
simulation, and the initial 20\% of the trajectory was discarded as a warm-up period. A linear regression was performed on the remaining data,
\begin{equation}
    E_{\mathrm{tot}}(t) = a_{\mathrm{tot}} t + b ,
\end{equation}
and the stability metric was defined as
\begin{equation}
    s = \left| \frac{a_{\mathrm{tot}}}{N_{\mathrm{atom}}} \right| ,
\end{equation}
which is reported in units of eV atom$^{-1}$ ps$^{-1}$, consistent with the LAMBench 
implementation. A trajectory with \(s < 5\times10^{-4}\) eV atom\(^{-1}\) ps\(^{-1}\) was considered to exhibit negligible
energy drift.
\begin{table*}[htbp]
    \centering
    \caption{
      Energy drift slopes from 50\,ps NVE molecular dynamics simulations on representative validation structures from OMol25, OMat24, and OC20M.
      For each dataset, the dense DPA3 baseline and MoE-E model are trained on the corresponding dataset.
      The MoE-E model uses 64 total experts, 3 shared experts, and \(K=6\) activated experts.
    }
    \label{tab:md_stability}
    \setlength{\tabcolsep}{5pt}
    \renewcommand{\arraystretch}{1.2}
    \small
    \begin{tabular}{llcc}
      \toprule
      \textbf{Dataset} & \textbf{Structure} & \textbf{Baseline} & \textbf{MoE-E} \\
      & & \multicolumn{2}{c}{Energy drift slope (eV atom$^{-1}$ ps$^{-1}$)} \\
      \midrule
      \multirow{3}{*}{OMol25}
      & \ce{NaAlOH}        & $2.41\times10^{-4}$ & $3.84\times10^{-5}$ \\
      & \ce{NSiNS2H}       & $1.88\times10^{-6}$ & $9.32\times10^{-5}$ \\
      & \ce{OPO2ClS2I}     & $3.78\times10^{-6}$ & $1.36\times10^{-6}$ \\
      \midrule
      \multirow{3}{*}{OMat24}
      & \ce{Pr2Si2W2}      & $1.35\times10^{-7}$ & $5.13\times10^{-7}$ \\
      & \ce{Ca4Y2W2}       & $4.39\times10^{-7}$ & $7.12\times10^{-7}$ \\
      & \ce{TePb6S5}       & $3.27\times10^{-7}$ & $1.10\times10^{-7}$ \\
      \midrule
      \multirow{3}{*}{OC20M}
      & \ce{Cs6CH2O}       & $1.76\times10^{-5}$ & $2.55\times10^{-5}$ \\
      & \ce{Te32Pd16C2H4O2} & $4.36\times10^{-6}$ & $4.94\times10^{-6}$ \\
      & \ce{H2Al32Cr48Mn16N2O} & $6.55\times10^{-3}$ & $1.38\times10^{-5}$ \\
      \bottomrule
    \end{tabular}
  \end{table*}

The energy drift slopes for all tested trajectories are summarized in 
Supplementary Table~\ref{tab:md_stability}. All MoE-E trajectories remain below the LAMBench stability threshold, indicating negligible
energy drift in the sparse expert model. The dense DPA3 baseline 
and MoE-E exhibit comparable stability on the OMol25 and OMat24 structures, 
with neither model showing a consistent advantage. On OC20M, the two
models are comparable for most structures, but the baseline trajectory 
for \ce{H2Al32Cr48Mn16N2O} exhibits a substantially larger energy drift than the corresponding MoE-E trajectory. Overall, the PES scans and
NVE simulations indicate that element-wise sparse routing does not introduce visible PES discontinuities or compromise MD stability.

\section{MoE on CHGNET}\label{sec:chgnet}

To assess whether the MoE design transfers to MLIP architectures beyond DPA3, 
we applied the element-wise MoE-E framework to CHGNET~\cite{deng2023chgnet}, a graph neural 
network potential that explicitly models atomic, bond, and angle features. CHGNET differs 
from DPA3 in its network topology and featurization, making it a suitable testbed for evaluating the generality of the MoE approach.

In CHGNET, the majority of parameters reside in the atom, bond, and angle 
convolutional layers and the associated MLPs. We replaced the dense MLP blocks in these 
layers with element-wise MoE-E layers. Each MoE-E layer contains 32 experts in total. 
We vary the number of activated experts $K=2,4,6$, with the number of shared experts 
set to $1,2,3$, respectively (i.e. half of the activated experts are shared). 
Because the full CHGNET dimensions (\texttt{atom\_fea\_dim}=64, \texttt{bond\_fea\_dim}=64, 
\texttt{angle\_fea\_dim}=64, with corresponding hidden dimensions of 64) led to 
excessive memory consumption when combined with the MoE expert pool, we 
reduced all feature and hidden dimensions by a factor of two: 
\texttt{atom\_fea\_dim}=32, \texttt{bond\_fea\_dim}=32, \texttt{angle\_fea\_dim}=32, 
\texttt{atom\_conv\_hidden\_dim}=32, \texttt{bond\_conv\_hidden\_dim}=32, 
\texttt{angle\_layer\_hidden\_dim}=16, and \texttt{mlp\_hidden\_dims}=[32, 32, 32].
 All other architectural hyperparameters were kept identical to the original CHGNET setting. 
 The comparison is therefore between a halved-dimension CHGNET baseline and halved-dimension CHGNET+MoE-E models.

The models were trained on the MPtrj dataset~\cite{mptrj} for 20 epochs. 
The best validation and test-set metrics are reported in Supplementary Table~\ref{tab:chgnet}.

\begin{table*}[htbp]
  \centering
  \caption{%
    Validation and test metrics for the CHGNET baseline and MoE-E extensions trained on MPtrj.
    Energy and force MAEs are reported in meV/atom and meV/\AA, respectively; stress MAE is reported in GPa.
    All MoE-E models use 32 total experts and halved feature dimensions (see text).
  }
  \label{tab:chgnet}
  \setlength{\tabcolsep}{10pt}
  \renewcommand{\arraystretch}{1.2}
  \begin{tabular}{l c c c c c c}
    \toprule
    \textbf{Model} & \textbf{Epoch} & \textbf{Shared experts} & \textbf{Val e} & \textbf{Test e} & \textbf{Test f} & \textbf{Test s} \\
    & & & (meV/atom) & (meV/atom) & (meV/\AA) & (GPa) \\
    \midrule
    Baseline small & 19 & -- & 37.1 & 35.9 & 80.8 & 0.3636 \\
    MoE topk=2   & 19 & 1 & 35.1 & 34.4 & 79.1 & 0.3612 \\
    MoE topk=4   & 19 & 2 & 34.3 & 33.1 & 77.1 & 0.3501 \\
    MoE topk=6   & 18 & 3 & 33.7 & 32.4 & 75.9 & 0.3468 \\
    \bottomrule
  \end{tabular}
\end{table*}

The MoE-E extension improves energy, force, and stress predictions relative to the halved-dimension CHGNET 
baseline on both the validation and test sets, with larger gains as the number of activated experts $K$ increases. 
These results indicate that the element-wise MoE design transfers to architectures beyond DPA3. 
The reduced dimensionality used here, necessitated by memory constraints, 
means that these results should be interpreted as a proof of concept for architectural 
transferability rather than a production-optimized benchmark. Optimizing the MoE 
implementation for full-scale CHGNET dimensions remains an important direction for future work.

\renewcommand{\refname}{Supplementary References}
\bibliography{supplement}